\documentclass[twocolumn,showpacs,showkeys,preprintnumbers,amsmath,amssymb]{revtex4}
\usepackage{graphicx}
\usepackage{dcolumn}
\usepackage{eucal}
\usepackage{amsfonts}
\usepackage{latexsym}
\usepackage{epsfig}
\usepackage{color}

\renewcommand{\d}{\mathrm{d}}

\DeclareMathSymbol{\mg}{\mathrel}{symbols}{"1D}
\newcommand{\ml}{\ll}

\newcommand{\dga}{{\dot{\alpha}}}
\newcommand{\dgb}{{\dot{\beta}}}

\newcommand{\ga}{\alpha}
\newcommand{\gb}{\beta}

\renewcommand{\ge}{\epsilon}

\newcommand{\gl}{\lambda}

\newcommand{\gth}{\theta}

\newcommand{\gs}{\sigma}

\newcommand{\gL}{\Lambda}

\newcommand{\gO}{\Omega}

\newcommand{\gPs}{\Psi}

\newcommand{\cD}{{\cal D}}

\newcommand{\cL}{{\cal L}}

\newcommand{\tF}{{\tilde F}}

\newcommand{\tV}{{\tilde V}}

\newcommand{\tr}{\text{tr}}

\newcommand{\ra}{\rightarrow}

\newcommand{\der}{\partial}

\newcommand{\labl}[1]{\label{#1}}
\newcommand{\Kh}{K\"{a}hler}
\newcommand{\beq}{\begin{equation}}
\newcommand{\eeq}{\end{equation}}
\newcommand{\barr}{\begin{array}}
\newcommand{\earr}{\end{array}}
\newcommand{\equ}[1]{\begin{gather} #1 \end{gather}}

\newcommand{\arry}[2]{\begin{array}{#1} #2 \end{array}}

\newcommand{\non}{\nonumber}
\newcounter{oldcounter}
\newcommand{\bD}{{\bar D}}
\newcommand{\bE}{{\bar E}}

\newcommand{\bS}{{\bar S}}

\newcommand{\bW}{{\bar W}}

\newcommand{\bgl}{{\bar\lambda}}

\newcommand{\bgth}{{\bar\theta}}
\newcommand{\bgs}{{\bar\sigma}}

\newcommand{\bgL}{{\bar\Lambda}}

\newcommand{\ea}{\end{array}\]}
\newcommand{\be}{\begin{equation}}
\newcommand{\ee}{\end{equation}}
\newcommand{\bea}{\begin{eqnarray}}
\newcommand{\eea}{\end{eqnarray}}

\newcommand{\mpl}{M_{Pl}}
\begin{document}
\preprint{UVIC-TH/04-05;~ FTPI-MINN-04/14;~ UMN-TH-2303/04;~ hep-ph/0404271}
\title{Lorentz Violation in Supersymmetric Field Theories}
\author {Stefan Groot Nibbelink$^{1}$
\ and
Maxim Pospelov$^{2}$
}
\affiliation{$^{1}$William I.\ Fine Theoretical Physics Institute,
University of Minnesota, Minneapolis, MN 55455, USA}
\affiliation{$^{2}$ Department of Physics and Astronomy,
University of Victoria, Victoria, BC, V8P 1A1, Canada.} 
\begin{abstract}\noindent
We construct supersymmetric Lorentz violating operators for matter and
gauge fields. We show that in the supersymmetric Standard Model the
lowest possible dimension for such operators is five, and therefore
they are suppressed by at least one power of an ultra--violet energy
scale, providing a possible explanation for the smallness of Lorentz
violation and its stability against radiative corrections. Supersymmetric 
Lorentz noninvariant operators do not lead to modifications of
dispersion relations at high energies thereby escaping constraints
from astrophysical searches for Lorentz violation.  
\end{abstract}
\maketitle
\vskip1pc

\section{Introduction}
\labl{sc:intro}

Recent years have seen an increase in the number of theoretical studies 
of Lorentz Violation (LV), as well as intensified experimental efforts 
searching for LV signatures in terrestrial, astrophysical and
cosmological settings \cite{Kost1,CG}. 
For example, {\em effective} LV at low energies may arise in 
string theory due to a non--vanishing background of an
anti--symmetric tensor field.
Alternative scenarios of  quantum gravity often predict that
at ultra--short distances particle dispersion relations are
modified by cubic and higher terms in the energy, (see, e.g. \cite{lcq}), 
\equ{ 
E^2 =p^2 +m^2 + b_1 \frac {E^3}M+ b_2 \frac{E^4}{M^2}\ldots, 
\label{conjecture}
}
where $b_i$ are some dimensionless constants. Although such conjectures
are undoubtedly very speculative, if true they could provide a
powerful tool of probing microscopic $M^{-1}$ distances via LV
physics.

LV operators can be classified according to their dimension. Cubic and
higher order modifications of dispersion relations correspond to  LV
operators of at least dimension five \cite{MP:}. According to naive
dimension counting, the dimension $D$ of an operator determines its
scaling $\sim M^{4-D}$ with the characteristic energy scale $M$ at
which the operator is generated. Hence, dimension five operators are
necessarily suppressed by one power of the ultraviolet scale
$M$. However even Planck mass ($\mpl$)--suppressed operators for
photons, electrons and quarks are ruled out  by a number of
astrophysical constraints and precision measurements up to  $\sim
10^{-5}$ level \cite{Vuc,GK,Ted1,MP:}. Even more serious problems 
arise with dimension three and four LV operators classified in \cite{CK}, 
since there are no dimensional arguments as
to why such operators should be small. Moreover, higher dimensional
operators will in general induce lower dimensional ones through loop
corrections with power--law divergent coefficients. Only
additional symmetry arguments may  provide genuine suppressions of
such lower dimensional operators \cite{MP:}.

An obvious candidate for such a symmetry is supersymmetry
(SUSY). Following the prevailing point of view in particle theory, we 
assume that at ultra--short distances (close to $1/\mpl$) SUSY is
realized  exactly. As the ultraviolet behavior of SUSY theories are
free of potentially dangerous quadratic divergences, it is generally
accepted as being a technical solution to the hierarchy problem. 
SUSY is conventionally introduced as a graded extension of the
Poincar\'e algebra generated by translations, rotations and Lorentz
transformations, therefore, one might expect that SUSY is
simply incompatible with LV physics. This is not the case because it
is possible to restrict all considerations to the subalgebra generated
by supercharges and translations only. In this work we only consider
LV SUSY theories that are representations of this algebra without any
further modifications. Moreover we only focus on the standard
chiral and vector superfields, which are conventionally used to
describe the field content of the Minimal Supersymmetric Standard
Model (MSSM). Of course, the constraints of Lorentz and rotational
invariance cannot be enforced anymore. However, we will see that SUSY
still provides a very powerful selection rule for LV
interactions. Moreover, like in conventional SUSY field theories we 
expect that operators forbidden by SUSY will be suppressed by
some power of $m_{\rm soft}/M$ below the soft SUSY breaking scale
$m_{\rm soft}$, leading to a possible partial explanation of why the
LV operators of dimension three are so tiny.

In this Letter we classify LV operators that are compatible with
exact SUSY for arbitrary vector and tensor backgrounds. 
To this end, we describe a systematic method of constructing
LV interactions in the SUSY context. We find that SUSY combined with
gauge invariance severely constraints the possible form of such operators.  
From this analysis we conclude that the smallest dimension of LV 
operators within the framework of the MSSM is five. We show that these
SUSY LV operators do not lead 
to significant modifications of  dispersion relations.

\section{supersymmetric LV Lagrangians}
\labl{sc:constr}

As stated earlier, LV preserves the subalgebra 
generated by supercharges and translations, thereby allowing the use of the
superspace technique. Even though this is equivalent to a component
approach, the superspace language permits the most straightforward and 
economical formulation of LV operators. To fix the notations we follow
the textbook by Wess \& Bagger \cite{WessBagger}. 
The matter and gauge fields in the MSSM are
described by chiral multiplets and vector multiplets. To facilitate
the counting the dimensions of LV operators from their superfield
expressions, we list the mass dimensions of objects appearing in this
Letter in the table below:  
\[
\arry{l|c|c|c|c|c|c|c|c}{
&&&&&&&&\\[-2ex] 
\text{obj.} & \der_m & \gth^\ga & D_\ga  & 
\int \d^2 \gth & \int \d^4 \gth & ~ S~  & ~ V~  & W_\ga 
\\[-2ex] &&&&&&&& 
\\\hline &&&&&&&&  \\[-2ex]
\text{dim.} & 1 & -\frac 12 & \frac 12 & 1 & 2 & 1 & 0 & \frac 32 
\\[-2ex] &&&&&&&&  
}
\]
Here $S$ denotes a chiral superfield, i.e.\ $\bD_\dga S = 0$, and 
$\bD_\dga$ is a super covariant derivative. The superfield strength   
$W_\ga = -\frac 14 \bD^2 \big( e^{-V} D_\ga e^V \big)$ is obtained
from the vector superfield $V$. With the use of this table, 
it follows  that
the standard Lagrangian for the Wess--Zumino model, 
\equ{
\cL_{WZ} = 
\int \d^2 \gth \, P(S) + \text{h.c.} + \int \d^4 \gth \, \bS S, 
\labl{WZst} 
}
with a (cubic) superpotential $P(S)$ has mass dimension four. 
Throughout this work we include the superspace measures in
the counting of the dimension of operators.

We construct SUSY LV operators coupled to background tensors that
lead to modifications of physical observables, like a preferred direction
or Lorentz frame. Our main result states that:  

{\em  
Any LV operator respecting MSSM gauge
invariance and exact SUSY has dimension five or higher 
and therefore is suppressed by at least one power of an 
ultraviolet scale $M$. 
}

We show this in three steps: First we classify LV operators for chiral
superfields, next we investigate consequences of gauge invariance, and
finally we apply our results to the MSSM. 
The fundamental chiral and vector multiplets, $S$ and $V$, do not
carry any Lorentz indices. As only the derivatives $D_\ga$, $\bD_\dga$
and $\der_m$ are SUSY preserving, SUSY LV interactions should be constructed 
by applying a number of these derivatives to superfields $S$ and $V$. 
Consequently, any SUSY LV interaction contains two or more superfields, 
otherwise it is a total derivative in superspace. This rules out a
LV generalization of the Fayet--Iliopoulos term $\int\d^4\gth\,V$.
The absence of external fermionic backgrounds implies that all SUSY LV
operators contain an even number of fermionic derivatives $D_\ga$ and
$\bD_\dga$. Combining these observations imply
that SUSY LV starts at dimension four. In particular, we find that 
possible LV operators for
chiral superfields (labeled by $a, b, c$) up to dimension five are
obtained as chiral integrals ($\int\d^2\gth$) of the superpotential terms
\equ{
S_a \der_m S_b, \quad 
S_a \der_m \der_n S_b, \quad 
S_a S_b \der_m S_c, 
\labl{SPLV}
}
and as full superspace integral ($\int\d^4\gth$) of 
\equ{
\bS_a \der_m S_b, 
\labl{KLV}
}
up to total derivatives in superspace. Of all these operators only the
first term in \eqref{SPLV} has dimension four; all others have dimension
five.

Next we proceed to LV in SUSY gauge theories. As $D_\ga$ and 
$\der_m$ break super gauge transformations $S \ra e^{-\gL} S$ and 
$e^V \ra e^{\bgL} e^V e^\gL$, we introduce covariant derivatives 
$\cD_\ga S  = e^{-V} D_\ga ( e^V S )$ and 
\equ{
\der_m S \rightarrow  \cD_m S = 
- \frac i4 \bgs^{\dga\ga}_m \cD_{\dga\ga} S = 
- \frac i4 \bgs^{\dga\ga}_m \bD_\dga \cD_\ga S.  
\labl{GaugeCovDer}
}
Contrary to $\der_m$, this covariant derivative does not respect 
chirality: 
\(
\bD_\dgb \cD_{\dga\ga} S = 2 \ge_{\dgb\dga} W_\ga S \neq 0, 
\labl{NonChiral}
\) 
and hence LV superpotentials \eqref{SPLV} cannot be generalized to
charged chiral superfields! Consequently, the only dimension five SUSY
LV operator for a charged chiral multiplet  is the gauge invariant
version of the \Kh\ LV term \eqref{KLV}:  
\equ{
\bS e^V \cD_m S. 
\labl{GKLV}
}
The constraints of gauge invariance for vector multiplets are similar
to standard Lorentz preserving theories, hence possible LV terms in the
SUSY gauge sector are the full superspace integral of  
\equ{
\tr\,  \bW_\dga  e^{V} W_\ga e^{-V}  
\labl{gaugeLV}
}
of dimension five and chiral integrals of 
\equ{
\tr\,  W_{(\ga} W_{\gb)}', 
\quad 
\tr\,  S W_{(\ga} W_{\gb)}', 
\quad 
\tr\,  W_\ga \der_m W_\gb',
\labl{gaugeCLV}
}
where the first expression has dimension four, while the other two
have dimension five.  The chiral superfield $S$ is in an adjoint
representation if $V$ is non--Abelian, and a gauge singlet for Abelian $V$. 
Where needed, we have preformed symmetrization of $\ga$ and $\gb$,
denoted by $(\ga,\gb)$, to project on  LV anti--symmetric tensor
background, $b^{mn} (\gs_{mn}\ge)^{\ga\gb}$ (which may appear in
non--commutative field theories, for example). For a single $U(1)$ or for
non--Abelian gauge multiplets the first term of \eqref{gaugeCLV}
vanishes. 

Now we apply these results to the MSSM: Since all MSSM chiral
superfields are charged under gauge symmetries, no LV superpotential is
allowed. In particular, a  LV generalization of the $\mu$ term 
in the Higgs sector, $H_1 \partial_m H_2$, is excluded by gauge
invariance. Since MSSM contains only one $U(1)$ vector multiplet, 
operator $\tr\,  W_{(\ga} W_{\gb)}'$ vanishes. Therefore all 
dimension four SUSY LV operators in the MSSM are excluded, and the LV 
terms start from dimension five. 
Moreover, not only are dimension four LV operators forbidden in the
MSSM, but also the number of dimension five operators is limited: The only
three types of operators are \Kh\ terms \eqref{GKLV} for MSSM chiral 
multiplets, interactions based on \eqref{gaugeLV} and the third
term in \eqref{gaugeCLV} for the MSSM vector multiplets.

Finally, we stress that in any SUSY theory LV is allowed only at
dimension four and higher. If the spectrum of MSSM at the electroweak
scale or below is extended by chiral singlets such as right--handed
neutrinos, and/or by additional $U(1)$ vector multiplet(s), dimension
four LV operators from (\ref{SPLV}) and (\ref{gaugeCLV}) can indeed
appear.

\section{Phenomenological consequences}

As shown above there exists only three possible types of dimension 
five LV operators that preserve SUSY in the MSSM. We investigate
phenomenological consequences of these operators and, in particular, 
we claim that: 

{\em
The dimension five SUSY LV operators do not lead to
significant modifications of dispersion relations. 
}

The physical reason for this result can be understood from the 
modification of the kinetic term \eqref{GKLV} for the scalar
component $z$ of a chiral superfield $S$. This modification 
$M^{-1}\bar z \partial_m \der^2z$ can be reduced on the equations of
motion to $M^{-1}m^2 \bar z \partial_m z$. The resulting  
$\sim M^{-1}m^2 E$ correction of the dispersion relation is small
w.r.t.\ $m^2$.

These arguments can be lifted to superspace. For simplicity, we focus
on LV in the Super Quantum Electrodynamics (SQED) 
part of the MSSM, as the extension to the full MSSM is
straightforward. The theory of SQED consists of a $U(1)$ vector
multiplet $V$ and two oppositely charged chiral superfields
$E_\pm$. The complete SQED Lagrangian with all dimension five 
SUSY LV terms is given by   
\equ{
\int \d^2 \gth\, \Big( \frac 1{16 e^2} W^2 + m E_+ E_- \Big) 
+ \text{h.c.} 
+ \int \d^4\gth\, \bE_\pm e^{\pm V} E_\pm 
\non 
\\ 
+ \frac 1M \int \d^4\gth\, \Big( 
i N_\pm^m\, \bE_\pm e^{\pm V} \cD_m E_\pm  
- \frac 12 N^m\, \bW \bgs_m W 
\Big) 
\non 
\\ 
+ \frac 1{M} \int \d^2 \gth\, C^{p\,mn}\, W \gs_{mn} \der_p W  
+ \text{h.c.}, 
\labl{LVSQED}
}
with $e$ the electric charge and $m$ the mass of the electron. The
first line gives the standard Lagrangian for SQED, while the other two
lines describe SUSY LV by external vectors $N^m_\pm$ and $N^m$, and a
tensor  $C^{p\,mn} = -C^{p\,nm}$.

To show that the dispersion relation of the electron/positron is not
significantly modified, we compute the superfield equations of
motion, 
\equ{
\der^2 E_\pm 
- m^2 \Big(1 - \frac iM ( N^m_\pm - N^m_\mp ) \der_m \Big) 
E_\pm = 0, 
}
up to first order in the LV and dropping all dependence on the vector superfield $V$. 
The resulting corrections to the dispersion relation,
\equ{
 E^2 = p^2+m^2 +{m^2} (N_\pm^0 - N_\mp^0) \frac {E}{M} + \ldots,
\label{drsusy}
}
are drastically {\em smaller} than the conjectured form
(\ref{conjecture}), and in fact, much smaller than $m^2$ as long as 
$E \ml M$!  
Further corrections with higher powers of $E$ are suppressed by additional
factors of $m/M$. The same holds for higher dimensional SUSY LV
operators, like $\bE_\pm e^{\pm V} \cD_m \cD_n \ldots E_\pm$. An even
stronger conclusion can be reached for the photon LV operators  
in \eqref{LVSQED}. The equation of motion in the presence of $N_m$
\equ{
\Big( 1 - N^m \bgs_m^{\dga\ga} \bD_\dga D_\ga \Big) 
D^\gb \bD^2 D_\gb \, V = 0. 
\label{photon}
}
can be solved iteratively to first order in LV parameter. The zeroth
order equation of motion can be applied in the second term 
of Eq. (\ref{photon}) after which it vanishes, leaving no modifications 
of photon propagation by $N_m$! Using similar approach, we can extend
this result to the $C^{pmn}$--proportional operator in \eqref{LVSQED}.

To understand some other phenomenological consequences of SUSY LV, 
we present the component form of $N^m$--proportional operator of 
\eqref{LVSQED}:  
\equ{
-\frac {N^m}{2M} \int \d^4 \gth \, \bW \bgs_m W = \frac {N^p}M  \Big[ 
\frac 12 \tilde F_{kp}\der_l F^{kl}- D \der^k F_{kp} 
\non \\[1ex] 
 \,+~\eta_{pk}\lambda \sigma^k \Box\,\overline{\lambda}\,  
	  -
	\, \lambda\, \sigma^m {\partial}_m\partial_p \, 
	   \overline{\lambda},
\Big], 
\label{d4thww}
}
where $F_{mn}$ is the electromagnetic field strength, $\lambda$ is the
photino, and $D$ is the auxilary field.
The spatial component of $N^m$ couples to the cross product of the 
electric field and the electric current, $({\bf E\times J})\cdot{\bf N}$ 
upon the replacement of $\der_l F^{kl}$ by the current $J^k$ in \eqref{d4thww}. 
Under descrete symmetries this interaction is CPT odd, P even, C and T odd. The average of 
${\bf E\times J}$ inside a particle with charged constituents, i.e. a
nucleon or a nucleus, is a vector directed along a nuclear spin ${\bf I}$. 
Following the method of \cite{MP:}, we estimate the size of an
effective interaction between ${\bf N}$ and ${\bf I}$ to be at the level of 
\(
H_{\rm eff}\sim (10^{-5}-10^{-3}) M^{-1}(1~{\rm GeV})^2  (\bf{N\cdot I}),
\label{pheno}
\)
where 1 GeV enters as a characteristic hadronic energy scale.
This is precisely the correlation searched for  by the clock
comparison experiments  (see e.g.\ \cite{PR} and references therein)
and ${\bf N}M^{-1}$ is limited typically at the level better than
$10^{-5}M_{Pl}^{-1}$.

\section{Lorentz Violating SUSY breaking}
\labl{sc:spurious}

LV operators constructed in Section \ref{sc:constr} respect SUSY
manifestly. Here we present a method to obtain LV Lagrangians that 
generically lead to SUSY breaking. Consider the Lagrangian
\equ{
\int \d^4 \gth\, \tV \, \gPs, 
\ \text{with}\ 
\tV = - n_m \gth \gs^m \bgth.
\labl{spuriousAc}
}
for an arbitrary (real composite) superfield $\gPs$. According to the
table in Section \ref{sc:constr} the superspace variables $\bgth_\dga$ and
$\gth_\ga$ in  $\tV$ effectively {\em reduce} the dimension of the 
operator by one. For example, by taking $\gPs = \bar SS$ we obtain
LV operators of dimension three. This construction does not preserve 
SUSY in general: Only if 
\equ{
\int \d^4 x\, \bD^2 D_\ga\gPs| = 0, 
\labl{IffSusy}
} 
the operator \eqref{spuriousAc} respects SUSY. (Here $|$ indicates
that $\gth_\ga$ and $\bgth_\dga$ are set to zero after all superspace
differentiations.)

This result can be used to show that LV by a Chern--Simons term, i.e.\ 
\(
n_m \ge^{mnkl} A_n \der_k A_l,
\)
does not have a SUSY extension. Using \eqref{spuriousAc}  
we obtain a Lagrangian that contains the Chern--Simons interaction
\equ{
M \! \int \d^4\gth\, \tilde V \gO = 
\frac {n_m}4 M  \Big( 
\ge^{mnkl} A_n \der_k A_l'   + \gl \gs^m \bgl' 
\Big), 
\labl{CSsusy}
}
where 
\(
\gO = - \frac 14 \big( 
D^\ga V W_\ga'  + \bD_\dga V \bW^{\prime \dga}  +  V D^\ga W_\ga'   
\big) 
\) 
is the Chern--Simons superfield \cite{Cecotti:1987nw}. 
(The construction works also for a single vector multiplet $V' = V$, 
and \eqref{CSsusy} is super gauge invariant because $\bD^2 D_\ga \tV=0$.) 
But the condition \eqref{IffSusy} implies that the Lagrangian
\eqref{CSsusy} does not respect SUSY, since  
\(
\bD^2 D_\ga \gO \big| = 
D_\ga \big( W^\gb  W_\gb' \big) \big|
\) 
does not vanish. Our conclusion supports the recent result of 
ref.\ \cite{Belich:} that Chern--Simons interactions require  
SUSY breaking. Moreover, by taking $V' = \der^m \der^n V$ in
\eqref{CSsusy},  we conclude that the only dimension five operator
that leads to $E^3$--modification of the photon dispersion relation
\cite{MP:}, 
\(
F^m{}_{p} \der^n \tF^{kp}  
\) 
breaks SUSY explicitly!

\section{Discussion and conclusion} 
\labl{sc:concl}

Before summarizing our main findings we comment on some recent
publications \cite{Berger:,Berger:2} that considered the construction 
of dimension three and four SUSY LV interactions for a (neutral) chiral
superfield, which seems to be in conflict with the main results of this work.
As the authors observe themselves, the dimension three operators can
be removed by a suitable (super)field redefinitions \cite{Berger:} leaving no 
observable effects. But they claim that the
modification of the Wess--Zumino  action by a symmetric tensor $k^{mn}$
combined with modified superalgebra and SUSY transformations gives
rise to viable LV effects. However, their resulting dimension four LV
Lagrangian (given in \cite{Berger:}) can be removed by the linear
change of coordinates,  $x_m' = x_m - k_{m}{}^{n}x_n$, which also
brings SUSY transformations to a usual Lorentz--conserving form.

We have presented a method of obtaining manifestly supersymmetric LV 
interactions by 
allowing free spacetime or an even number of super covariant
derivatives to act on superfield expressions. We proved that exact 
SUSY requires LV to start at dimension four or higher. 
Gauge invariance and chirality prohibits derivatives on charged chiral 
superfields to appear in the superpotential. Therefore LV in the charged
chiral multiplet sector begins at dimension five since extra derivatives are 
allowed only in kinetic terms.  Applying our results to the
superfield content of the MSSM we arrive at our central conclusion:
All possible LV operators in MSSM have at least dimension five and
therefore are suppressed by one or more powers of a large
ultra--violet scale responsible for LV. Dimension five SUSY LV
interactions for SQED are given in (\ref{LVSQED}) with obvious 
generalization to full MSSM. 

None of the SUSY LV operators lead to significant high--energy
modifications of the dispersion relations. We find that SUSY LV
operators can be reduced on the equations of motion,
producing an additional suppression by $m^2$ and 
suggesting a generic form of the SUSY LV dispersion relation:  
\equ{
E^2 =p^2 +m^2\left(1 + b_1 \frac {E}M+ b_2 \frac{E^2}{M^2}+\ldots\right), 
\label{whatitturnsouttobe}
}
which is in sharp contrast with (\ref{conjecture}), and does not 
modify propagation of photons. Therefore, SUSY LV leaves no imprint on
the propagation of high--energy particles and escapes constraints from
astrophysical searches of LV, but  can be probed with precision
measurements at low energies.

As exact SUSY forbids dimension three LV operators the problem of
dimensional transmutation of dimension five LV operators to dimension
three with quadratically divergent loop coefficients is solved. In a
more realistic theory SUSY needs to be broken, and dimensions three
operators may be generated but the quadratic  loop divergencies 
are stabilized at the soft breaking scale. 
Details of SUSY LV phenomenology with inclusion of soft
SUSY breaking and loop corrections will be investigated elsewhere
\cite{BGNP}.

We would like to thank P.\ Bolokhov, C.\ Burgess and R.\ Myers for useful discussions. 
The work of SGN is supported in part by the Department of Energy under
contract DE--FG02--94ER40823 at  the University of Minnesota.  The
research of MP is supported in part by NSERC of Canada.

\bibliographystyle{apsrev}
\bibliography{lorentz}

\end{document}